%This is a plain TEX file.
%All macros are listed below.
%%%%%%%%%%%%%%%%%%%%%%%%%%%%%%%%%%%%%%
\def\fr#1/#2{{\textstyle{#1\over#2}}} 
\newcount\ftnumber
\def\ft#1{\global\advance\ftnumber by 1
          {\baselineskip 12pt    
          \footnote{$^{\the\ftnumber}$}{#1}}}
%%%%%%%%%%%%%%%%%%%%%%%%%%%%%%%%%%%%%%%%%%%%%%%

\newcount\eqnumber
\def\eq(#1){
    \ifx\DRAFT\undefined\def\DRAFT{0}\fi	%if undef'd, make it 0
    \global\advance\eqnumber by 1%
    \expandafter\xdef\csname !#1\endcsname{\the\eqnumber}%
    \ifnum\number\DRAFT>0%
	\setbox0=\hbox{#1}%
	\wd0=0pt%
	\eqno({\offinterlineskip
	  \vtop{\hbox{\the\eqnumber}\vskip1.5pt\box0}})%
    \else%
	\eqno(\the\eqnumber)%
    \fi%
}
\def\(#1){(\csname !#1\endcsname)}

%%%%%%%%%%%%%%%%%%%%%%%%%%%%%%%%%%%%
\baselineskip=15pt
\magnification=\magstep1
%%%%%%%%%%%%%%%%%%%%%%%%%%%%%%%%%%
\def\zv{{\bf z}}
\def\xv{{\bf x}}
\def\al{\alpha}
\def\be{\beta}
\def\s{{\cal S}}
\def\h{{\cal H}}
\def\>{\rangle}
\def\<{\langle}
\def\u{\uparrow}
\def\d{\downarrow} 

\def\k#1{|#1\>}
\def\b#1{\<#1|}
\def\ks{\k\phi}
\def\bs{\b\phi}

\def\h{{\cal H}}
\def\x{\otimes}
\def\tr{{\rm Tr}}
\def\h{\fr1/2}

\def\+{\oplus}
\def\tr{{\rm tr}}
\def \ip#1#2{\langle #1 | #2 \rangle}
\def \imp{\Longrightarrow}

\centerline{{\bf THE ITHACA INTERPRETATION OF QUANTUM MECHANICS}}
\bigskip \centerline{N. David Mermin} \medskip \centerline{Cornell
University, Ithaca, New York, U.S.A.}  \bigskip 

{
\baselineskip=12pt 
\centerline{{\it Notes for a lecture given at the
 Golden Jubilee Workshop on Foundations}}
\centerline{{\it  of Quantum
Theory,  Tata Institute, Bombay, September 9-12,
1996}}} 

\bigskip
{\narrower \narrower \baselineskip = 12pt I list several strong
requirements for what I would consider a sensible interpretation of
quantum mechanics and I discuss two simple theorems.  One, as far as I
know, is new; the other was only noted a few years ago.  Both have
important implications for such a sensible interpretation.  My talk
will not clear everything up; indeed, you may conclude that it has not
cleared anything up.  But I hope it will provide a different
perspective from which to view some old and vexing puzzles (or, if you
believe nothing needs to be cleared up, some ancient verities.)\par}

\bigskip
\centerline{{\bf I. Introduction:  A Strategy for Constructing an
Interpretation.}}
\medskip

I'd like to describe some thoughts about what ought to go into a
satisfactory interpretation of quantum mechanics.  I do this with
considerable trepidation.  ``Ought to'' can be a highly personal
business.  And I have yet to put all the pieces together in a fully
convincing way.  Those who feel they understand quantum mechanics may
find what I have to say boring and self--indulgent, while those who
are bothered by quantum mechanics may find what follows inadequate or
even self--contradictory.  So you may get nothing out of my talk
beyond a description of two elementary theorems.  And one, and perhaps
even both of the theorems may be already known to you.  

I offer this half baked concoction nevertheless because
it seems to me the implications of the theorems for the interpretation
of quantum mechanics have not been emphasized and deserve some serious
exploration.  I've been thinking about them on and off for about half
a year now, and have found, to my surprise, that they keep resonating
in illuminating ways with various aspects of the Copenhagen
interpretation that have always struck me as anthropomorphic or
obscure.  I have been getting sporadic flashes of feeling that I may
actually be starting to understand what Bohr was talking about.
Sometimes the sensation persists for many minutes.  It's a little like a
religious experience and what really worries me is that if I am on the
right track, then one of these days, perhaps quite soon, the whole
business will suddenly become obvious to me, and from then on I will
know that Bohr was right but be unable to explain why to anybody
else.  

So it's crucial that I try to communicate some of these ideas before
they become so clear to me that only I can understand them.  The
problem, of course, is that my fragmentary vision may
be more of a pipe dream than a religious experience --- not a satori
but a bad trip.  I shall take that risk, and I ask for your indulgence.

I have a simple strategy for constructing an interpretation of quantum
mechanics: First of all, by ``quantum mechanics'' I mean quantum
mechanics as it is --- not some other theory in which the time
evolution is modified by non-linear or stochastic terms, nor even the
old theory augmented with some new physical entities (like Bohmian
particles) which supplement the conventional formalism without
altering any of its observable predictions.  I have in mind ordinary
everyday quantum mechanics.

I myself have never met an interpretation of quantum mechanics
I didn't dislike.  I shall try to extract something constructive from
all these strongly held negative intuitions, by prohibiting from my own
interpretation all of the features I have found unreasonable in all
the various interpretations I have encountered.  These prohibitions
are listed as the first five desiderata below.

To live with so many requirements I need room for maneuver.  This is
provided by adopting, as my sixth and final desideratum, the view that
probabilities are objective intrinsic properties of individual
physical systems.  I freely admit that I cannot give a clear and
coherent statement of what this means.  The point of my game is
to see if I can solve the interpretive puzzles of quantum mechanics,
{\it given\/} a primitive, uninterpreted notion of objective probability.  If
all quantum puzzles can indeed be reduced to the single puzzle of
interpreting objective probabilities, I would count that as progress.
Indeed since it is only through quantum mechanics that we have
acquired any experience of intrinsically probabilistic phenomena, it
seems to me highly unlikely that we {\it can\/} make sense of
objective probability without first constructing a clear and coherent
formulation of quantum mechanics in terms of such probabilities.

\bigskip
\centerline{{\bf II. Six Desiderata for an Interpretation of Quantum
Mechanics.}}
\medskip

Here are my own personal desiderata for a satisfactory
interpretation.  Most are based on my persistent discomfort with
various commonly held claims about the nature of quantum mechanics.  

\bigskip
\noindent{\bf (1) The theory should describe an objective reality
independent of observers and their knowledge.} 
\medskip

The maddening thing about the wave--function is the way in which it
manages to mix up objective reality and human knowledge.  As a clear
indication of this murkiness note that even today there is coexistence
between those who maintain that the wave--function is entirely real
and objective --- notably advocates of Bohmian mechanics or seekers of
a modified quantum mechanics in which wave--function collapse is a
ubiquitous real physical phenomenon---and those who maintain,
unambiguously with Heisenberg and presumably with Bohr, that the
wave--function is nothing more than a concise encapsulation of our
knowledge.  

A satisfactory interpretation should be unambiguous about what has
objective reality and what does not, and what is objectively real
should be cleanly separated from what is ``known''.  Indeed, knowledge
should not enter at a fundamental level at all.

\bigskip
\noindent{\bf (2) The concept of measurement should play no
fundamental role.} 
\medskip 

I agree with John Bell.\ft{{\it Against `measurement'\/}, Physics
World, 33-40, August, 1990.} There is a world out there,
whether or not we choose to poke at it, and it ought to be possible to
make unambiguous statements about the character of that world that
make no reference to such probes.  A satisfactory interpretation of
quantum mechanics ought to make it clear why ``measurement'' keeps
getting in the way of straight talk about the natural world;
``measurement'' ought not to be a part of that straight talk.
Measurement should acquire meaning from the theory --- not
vice--versa.

The view that physics can offer nothing more than an algorithm telling
you how to get from a state preparation to the results of a measurement
seems to me absurdly anthropocentric; so does limiting what we can
observe to what we can produce (``state preparation'' being one of the
things you can do with a ``measurement apparatus'').  Physics ought to
describe the unobserved unprepared world.  ``We'' shouldn't have to be
there at all.

\bigskip
\noindent{\bf (3) The theory should describe individual systems --- not
just ensembles.}
\medskip

The theory should describe individual systems because the world
contains individual systems (and is one itself!) and the theory ought
to describe the world and its subsystems.  Two attitudes lurk behind
every ensemble interpretation.  The first is a yearning (not always
acknowledged) for hidden variables.  For the notion that probabilistic
theories must be about ensembles implicitly assumes that probability
is about ignorance.  (The ``hidden variables'' are whatever it is that
we are ignorant of.)  But in a non-determinstic world probability has
nothing to do with incomplete knowledge, and ought not to require an
ensemble of systems for its interpretation.

The second motivation for an ensemble interpretation is the intuition
that because quantum mechanics is inherently probabilistic, it only
needs to make sense as a theory of ensembles.  Whether or not
probabilities can be given a sensible meaning for individual systems,
this motivation is not compelling.  For a theory ought to be able to
{\it describe\/} as well as {\it predict\/} the behavior of the world.
The fact that physics cannot make deterministic {\it predictions\/}
about individual systems does not excuse us from pursuing the goal of
being able to {\it describe\/} them as they currently are.  

\bigskip \noindent {\bf (4) The theory should describe small isolated
systems without having to invoke interactions with anything external.} 
\medskip 

Not only should the theory describe individual systems, but it should
be capable of describing {\it small\/} individual systems.  We apply
quantum mechanics all the time to toy universes having state--spaces
of only a few dimensions.  I would like not only to be able to do
that, as I now can, but to understand what I am talking about when I
do it, as I now cannot.

In particular I would like to have a quantum mechanics that does not
require the existence of a ``classical domain''.  Nor should it rely
on quantum gravity, or radiation escaping to infinity, or interactions
with an external environment for its {\it conceptual\/} validity.
These complications may be important for the practical matter of
explaining why certain probabilities one expects to be tiny {\it
are\/}, in fact tiny.  But it ought to be possible to deal with high
precision and no conceptual murkiness with small parts of the universe
if they are to high precision, isolated from the rest.

\bigskip \noindent {\bf (5) Objectively real internal properties of an
isolated individual system should not change when something is done to
another non-interacting system.} \medskip 

I agree with Einstein:\ft{{\it Albert Einstein:
Philosopher-Scientist\/}, ed. P. A. Schillp, Open Court, La Salle,
Illinois, 1970, p.~85.} ``On one supposition we should, in my opinion,
absolutely hold fast: the real factual situation of the system $\s_2$
is independent of what is done with the system $\s_1$, which is
spatially separated from the former.'' Indeed, I would take take
spatial separation to be just a particularly clear--cut way of
establishing the absence of mediating interactions between the two
systems, and apply the supposition --- generalized Einstein locality
--- to any two non-interacting systems.  

Einstein used his supposition, together with his intuitions about what
constituted a real factual situation, to conclude that quantum
mechanics offers an incomplete description of physical reality.  I
propose to explore the converse approach: assume that quantum
mechanics does provide a complete description of physical reality,
insist on generalized Einstein--locality, and see how this constrains
what can be considered physically real.

\bigskip \noindent {\bf (6) It suffices {\it (for now)\/} to base the
interpretation of quantum mechanics on the {\it (yet to be
supplied)\/} interpretation of objective probability.} 
\medskip

I am willing at least provisionally to base an interpretation of
quantum mechanics on primitive intuitions about the meaning of
probability in individual systems.

Quantum mechanics has taught us that probability is more than just a
way of dealing systematically with our own ignorance, but a
fundamental feature of the physical world.  But we do not yet
understand objective probability.  Popper\ft{{\it Quantum Theory and
the Schism in Physics,\/} Rowman and Littlefield, Totowa, New Jersey,
1982.} insisted that we cannot think correctly about quantum mechanics
until we learn how to think correctly about probability as an
objective feature of the world --- that the interpretation of quantum
mechanics had never squarely faced this issue.  I think he was right
about that, but wrong in maintaining that with his own formulation of
objective probability he had cleared up the conceptual puzzles.  

I don't have an understanding of objective probability any better than
Popper's, but I maintain that if we can make sense of quantum mechanics
conditional upon making sense of probability as an objective property
of an individual system, then we will have got somewhere.  Indeed, I
doubt that we can hope to understand objective probability until we
have achieved the partial success of making sense of quantum
mechanics, modulo such an understanding.  Quantum mechanics is our
only source of clues about what objective probability might mean, and
we will only unearth those clues if we can succeed in making sense of
quantum mechanics from such a perspective.  

So my attitude is this:  Assume that some wise person has come up with
an acceptable notion of probabilities as objective properties of
individual systems, and see if one can sweep all the puzzles of
quantum mechanics --- what Popper called the muddle, mysteries, and
horrors --- under that single accommodating rug.

\bigskip In summary, these are my Six Desiderata for an interpretation
of quantum mechanics:
\medskip

{\sl

\leftline{(1) Is unambiguous about objective reality.}

\leftline{(2) Uses no prior concept of measurement.}

\leftline{(3) Applies to individual systems. }

\leftline{(4) Applies to (small) isolated systems.}

\leftline{(5) Satisfies generalized Einstein--locality.}

\leftline{(6) Rests on prior concept of objective probability.}}

\bigskip
    
To persuade you that my aspirations are not made entirely of fluff,
let me next digress to tell you about two elementary theorems of
quantum mechanics that seem only recently to have been noticed.
 
\bigskip
\centerline{{\bf III. Two Elementary Theorems}}
\medskip

I shall describe in a naive way two elementary theorems of quantum
mechanics, which bear on the interpretive problem.  By ``naive'' I
mean that I shall use uncritically terms forbidden by Desideratum (2)
like ``measurement'', ``results of a measurement'', etc., because they
are a code we all understand, and because avoiding them would make the
purely mathematical argument much more clumsy.  I shall return to more
careful talk when I discuss the relevance of these theorems for the
interpretation of quantum mechanics.

To motivate the first theorem, consider the simplest possible quantum
mechanical system: a single two-state system, represented as the spin
of a spin-$\fr1/2$ particle.  Let this system be described by the
density matrix $$W = \h\k{\u_z}\b{\u_z} + \h\k{\d_z}\b{\d_z}.\eq(Wz)$$
This density matrix has many alternative representations. among them
being $$W = \h\k{\u_x}\b{\u_x} + \h\k{\d_x}\b{\d_x}.\eq(Wx)$$ The
first form is usually said to describe a situation in which the system
is in the state $\k{\u_z}$ with probability $\fr1/2$ and in the state
$\k{\d_z}$ with probability $\fr1/2$; the second, a situation in which
the equally probable states are $\k{\u_x}$ and $\k{\d_x}$.

{\it Is there an objective difference between these two situations?\/}
The statistics of all possible measurements one can make are, of
course, the same in both cases because the density matrix is the same,
but is there nevertheless an objective difference between a spin with
a definite but random polarization along $z$ and a definite but random
polarization along $x$?

There is no agreement on this elementary conceptual point.  People who
take the quantum state to be an objective property of an
individual system would say there is a difference: in one case this
objective property is unknown, but is equally likely to be $\k{\u_z}$
or $\k{\d_z}$; in the other case it is either $\k{\u_x}$ or
$\k{\d_x}$.  

But if you accept Desideratum (5) there can be no objective
difference.  For one can introduce a second two--state system that
does not currently interact with the first, taking the two systems to
be in the singlet state $$\k\Psi = \fr1/{\sqrt2}\k{\u_z}\k{\d_z} -
\fr1/{\sqrt2} \k{\d_z}\k{\u_z}\eq(Psiz)$$ which can equally well be
written $$\k\Psi = \fr1/{\sqrt2}\k{\u_x}\k{\d_x} - \fr1/{\sqrt2}
\k{\d_x}\k{\u_x}. \eq(Psix)$$ The representation \(Psiz) of $\k\Psi$
establishes that one can produce the situation suggested by the
representation \(Wz) of $W$ by measuring $\sigma_z$ on the
non-interacting ancillary system, while the representation \(Psix)
establishes that one can produce the situation suggested by \(Wx) by
measuring $\sigma_x$ on the ancilla.  If objectively real internal
properties of an isolated individual system are not to depend on what
is done to another non-interacting system, then there can be no
difference between these two realizations of the density matrix $W$.

This is the position of those who maintain that
Einstein--Podolsky--Rosen correlations and Bell's Theorem establish
only that there can be no local hidden-variables underlying quantum
mechanics, but do not establish that quantum mechanics itself implies
non-locality.  I would like to explore where one can get by adhering
to this view.

I once thought this peculiar situation --- the ability remotely to
produce either of two apparently distinct realizations of the same
density matrix $W$ --- stemmed from the degeneracy of $W$.  But this
is wrong.  Consider, for example, the non-degenerate density matrix $$W =
p\k{\u_z}\b{\u_z} + q\k{\d_z}\b{\d_z}\eq(Wpq)$$ with $p \neq q$, which
in spite of its non-degeneracy also has many alternative
representations, one of which is $$W =
\h\k{R}\b{R} + \h\k{L}\b{L},\eq(WRL)$$ where and $\k R$ and $\k L$
are the (non-orthogonal) states $$\k R = \sqrt p\k{\u_z} + \sqrt
q\k{\d_z}\eq(R)$$ and $$\k L = \sqrt p\k{\u_z} - \sqrt
q\k{\d_z}.\eq(L)$$ To make talking about things simple suppose that the
probability $p$ is very much larger than the probability $q = 1-p$.
Then interpretation \(Wpq) of the density matrix describes a system
that is in the state $\k{\u_z}$ with high probability and in the state
$\k{\d_z}$ with low probability, while the interpretation \(WRL)
describes a system that is with equal probability in one of two
non-orthogonal states representing spin along an axis tilted just
slightly away from \zv\ in either the direction \xv\ or $-\xv$.  

Again one can ask whether there is an objective difference between
these two apparently quite different situations, and again the answer
must be no.  For one can now introduce a second non-interacting
two-state system with the pair in the state $$\k\Psi =
\sqrt p\k{\u_z}\k{\u_z} + \sqrt q \k{\d_z}\k{\d_z} \eq(Ppq)$$ which can
equally well be written $$\k\Psi = \fr1/{\sqrt2}\k{R}\k{\u_x}+
\fr1/{\sqrt2}
\k{L}\k{\d_x}, \eq(Pab)$$   since $\k{\u_x}$ and $\k{\d_x}$ are
explicitly
$$\k{\u_x} = \fr1/{\sqrt2}\k{\u_z} + \fr1/{\sqrt2}\k{\d_z},$$
$$\k{\d_x}  = \fr1/{\sqrt2}\k{\u_z} - \fr1/{\sqrt2}\k{\d_z}.\eq(Px) $$
One can produce the situation associated with the
representation \(Wpq) of $W$ by measuring $\sigma_z$ on the
non-interacting ancilla, while one can produce the situation suggested by \(WRL) by
measuring $\sigma_x$ on the ancilla.

It is the content of Theorem I that this state of affairs is completely
general:\ft{N. Gisin, Helv. Phys. Acta {\bf 62}, 363 (1989)}$^,$\ft{L.
P. Hughston, R. Jozsa, and W. K. Wootters, Phys. Lett. A {\bf 183}, 14
(1993).}$^,$\ft{See Appendix A for a proof that is more complete than
Gisin's, and conceptually more straightforward than that of Hughston
et al.} 

\leftline {{\bf Theorem I:}}
\nobreak
Given an arbitrary system described by a $d$-dimensional density
matrix $W$, and given $N$ different interpretations of that density
matrix in terms of ensembles of systems in different (not-necessarily
orthogonal) pure states, associated with the expansions $$ W =
\sum_{\mu=1}^{D_n} p_\mu^{(n)}\k{\phi_\mu^{(n)}}\b{\phi_\mu^{(n)}},
\ \ \ \ n = 1,2,\ldots N,\eq(Wn)$$ then if $D$ is the largest of the
$D_n$ there is a state $\k\Psi$ in $d\times D$ dimensions and $N$
different observables $A_n$ in the $D$ dimensional ancillary subspace
such that measuring the observable $A_n$ on the ancilla leaves the
original $d$-dimensional subsystem in the state $\k{\phi_\mu^{(n)}}$
with probably $p_\mu^{(n)}$. 

If you take Desideratum (5) seriously, then there can be no more
objective reality to the different possible realizations of a density
matrix, then there is to the different possible ways of expanding a
pure state in terms of different complete orthonormal sets.  This is
not to say that the ``ignorance interpretation'' of a density matrix
does not provide a useful technical way to deal with ensembles of
systems.  But in the case of an individual system the density matrix
must be a fundamental and irreducible objective property, whether or
not it is a pure state.

The case of EPR correlations has made familiar the fact that when a
system is in a pure state that is not a simple product over
subsystems, then its subsystems can have no pure states of their own.
As far as I can tell, however, there is no consensus on whether to
take the subsystem density matrices as complete objective
characterizations of their internal properties.  In view of Theorem I,
Desideratum (5) requires us to do so.\ft{Note that this same
requirement, in a rather different context, alters the character of
the ``quantum measurement problem'': if a pure state for the
system--apparatus supersystem is entirely compatible with density
matrices for each subsystem, then the von Neumann ``collapse'' in a
measurement is not from a pure state to a mixture, but from viewing
the subsystem density matrices as fundamental and irreducible, to
viewing them under the conventional ignorance interpretation.}

\bigskip

The second theorem also applies to EPR correlations, but will be used
here in a much more general context.  To motivate it consider two
spin-$\fr1/2$ particles in the singlet state $\ks$.  Famously, their
spin components are perfectly anti-correlated.  In particular $$ \bs
\sigma_\mu^{(1)} \sigma_\mu^{(2)} \ks = -1,\ \ \
\mu=x,y,z.\eq(singlet)$$ There is a (less famous) coverse of
\(singlet):

If a system consisting of two spin-$\fr1/2$ particles has a density
matrix $W$, and if $$ \tr\, W \sigma_\mu^{(1)} \sigma_\mu^{(2)} = -1,\
\ \ \mu=x,y,z,\eq(converse)$$ then $W$ is necessarily the projection
operator on the singlet state: $$W = W_0 = \ks\bs = {1-
\sigma^{(1)}\cdot \sigma^{(2)} \over 4} .\eq(W0)$$ This is a direct
consequence of the fact that $W = W_0$ if and only if $\bs W \ks = 1$,
but if $W$ satisfies \(converse) then $$ \bs W \ks = \tr\, WW_0 =
\tr\, W\Bigl({1- \sigma^{(1)}\cdot \sigma^{(2)}
          \over 4}\Bigr) = \fr1/4\bigl(1 - \sum_{\mu=x,y,z}\tr\, W
           \sigma_\mu^{(1)} \sigma_\mu^{(2)}\bigr) = 1.\eq(QED)$$

There is a way of looking at this trivial result that makes it a
little surprising.  Suppose you have an ensemble of pairs of
spin$\fr1/2$ particles and you want to know if they all have total
spin zero.  Total spin being a global property of the pair, one way to
determine this would be to measure the total spin of enough pairs to
convince yourself that you are always going to get the result 0.  But
suppose the pairs are so far apart that this is impractical.  There is
another way.  Two people can do a series of separate measurements of
the two $x$ components to convince themselves that they are always
anti-correlated, and then do the same for the $y$ and $z$ components.
In this way they can establish a global property of an entangled state
by a series of local measurements together with the exchange of
information about the results of those local measurements.

It is the content of Theorem II that this intriguing state of affairs
is entirely general:\ft{N. D. Mermin, Cornell lecture notes
(unpublished), 1995. This must have been noticed before, but I have
not yet unearthed it in the literature.}

\leftline {{\bf Theorem II:}}
\nobreak
Given a system $\s = \s_1 \+ \s_2$ with density matrix $W$, then $W$
is completely determined by the values of tr$W\,A\x B$ for an
appropriate set of observable pairs $A$, $B$, where $A = A\x 1$ is an
observable of subsystem $\s_1$ and  $B = 1\x B$ is an
observable of subsystem $\s_2$. The proof is as follows:\ft{I give the
argument only for finite dimensional state spaces, leaving the
extension to the infinite dimensional case to those more
mathematically knowledgeable than I am.}

Let the $M_i$ be a set of hermitian operators that form a basis for the
algebra of operators on the subsystem $\s_1$ and let the $N_i$ be a
similar set for $\s_2$.  (If the state space for $\s_1$ is given an
orthonormal basis of states $\k{\psi_\mu}$ then the $M_i$ could, for
example consist of all the operators $\k{\psi_\mu}\b{\psi_\nu} +
\k{\psi_\nu}\b{\psi_\mu}$ and all the operators
$i(\k{\psi_\mu}\b{\psi_\nu} -
 \k{\psi_\nu}\b{\psi_\mu}$.)  Since the set of all 
$M_i\x N_j$ is a basis of hermitian operators for the algebra of
operators on the full system $\s$, it follows that if $\k\phi$ is any
state of $\s$ then the projection operator on $\phi$ has an expansion
of the form
$$\k\phi\b\phi = \sum_{i,j} c_{ij}(\phi) M_i\x N_j,\eq(expansion)$$
where the coefficients $c_{ij}$ are (real) numbers that can be
explicitly calculated for any state $\k\phi$ and any choice of the 
sets of operators $M_i$ and $N_i$.  So if $W$ is the density matrix
of $\s$ then 
$$\b\phi W \k\phi = \sum_{i,j} c_{ij}(\phi)\tr\, W\,M_i\x
N_j.\eq(pWp)$$ 

Therefore one can determine any diagonal matrix element of the density
matrix $W$ of an ensemble of systems $\s = \s_1\+\s_2$ from the
correlations in the results of an appropriate series of measurements
of observables specific to the subsystems $\s_1$ and $\s_2$.  Since an
arbitrary off-diagonal matrix element can be expressed in terms of
diagonal ones, $$\b\be W \k\al = \fr1/2 \b{\al+\be}W\k{\al+\be}
                + \fr i/2 \b{\al+i\be}W\k{\al+i\be}
                -\fr{1+i}/2\bigl(\b\al W \k\al + \b\be W \k\be\bigr),
\eq(arb)$$ we can determine in this way all the matrix elements of
the density matrix $W$ in some complete orthonormal basis for $\s$,
and hence determine $W$ itself.

This proof easily generalizes to a system $\s = \s_1\+\cdots\+\s_n$
composed of more than two subsystems: given any resolution of $\s$
into $n$ subsystems, the density matrix of $\s$ is entirely determined
by the correlations among appropriate observables belonging to those
subsystems.  In such cases the structure of quantum mechanics
guarantees the important fact that it doesn't matter whether we pin
down the density matrix, for example, of $\s = \s_1 \+ \s_2 \+ \s_3$
from correlations between observables of $\s_1$ with observables that
act globally on $\s_2\+\s_3$, or from correlations between observables
of $\s_3$ with observables acting globally on $\s_1\+\s_2$, or from
tripartite correlations between observables acting only on the three
subsystems.

Thus the density matrix of a composite system determines all the
correlations among the subsystems that make it up and, conversely,
{\it the correlations among all the subsystems completely determine
the density matrix for the composite system they make up.\/} The
mathematical structure of quantum mechanics imposes constraints, of
course, on what those correlations can be --- namely they are
restricted to those that can arise from some global density
matrix.\ft{That they cannot be more general than that is the content
of Gleason's Theorem.  It would be interesting to explore the extent
to which the underlying structure of probabilities assigned to
subspaces of a Hilbert space on which Gleason's Theorem rests is
itself pinned down by the requirement of consistency among different
possible resolutions of a system into subsystems.} The particular form
of that density matrix is then completely pinned down by the
correlations themselves.

This is familiar in the case $n=1$, where it reduces to the
fact that the set of all mean values over the entire system determines
the density matrix.  What seems to have been overlooked, and what
Theorem II establishes is the additional fact that for {\it any\/}
resolution of $\s$ into non-trivial subsystems $\s_1,\ldots,\s_n$, it
suffices to
determine $W$ to know those mean values only for a set of
observables restricted to those of the form $A_1
\x\cdots\x A_n$ where $A_j$ acts only on $\s_j$.
\bigskip

In the context of the Six Desiderata, Theorem I asserts that the
fundamental irreducible objective character of an individual system is
entirely specified by its density matrix, and Theorem II then tells us
that {\it the fundamental irreducible objective character of an
individual system is entirely specified by all the correlations among
any particular set of the subsystems into which it can be
decomposed.\/}

\bigskip
\centerline{{\bf IV.  The Ithaca Interpretation of Quantum Mechanics}}
\medskip

Having only begun looking at quantum mechanics from the point of view
of my six Desiderata and two Theorems, I have only
scattered, incomplete conclusions to report.    At this stage the
Ithaca Interpretation is rather fragmentary.  Central to it is the
doctrine that {\it the only proper subjects of physics are
correlations among different parts of the physical world.\/}
Correlations are fundamental, irreducible, and objective.  They
constitute the full content of physical reality.  There is no absolute
state of being; there are only correlations between subsystems.

Once it occurs to you to put it this way it sounds like a trivial point.
For how could it be otherwise?  One might imagine a God existing
outside of the World with direct unfathomable Access to its Genuine
Essence.  But physics is more modest in its scope than theology.  It
aims to understand the world in the world's own terms, and therefore
aims only to relate some parts of the world to others.  For
physicists, if not for theologians, this reduction in scope ought not
to be a serious limitation.

If correlations are the fundamental, irreducible, objective components
of physical reality, and physical reality consists of individual
systems, then probabilities are fundamental, irreducible, objective
properties of individual systems.  For among the possible correlations
among subsystems are those between projection operators associated
with the subsystems, which have an immediate interpretation as joint
probability distributions.  This raises difficult questions about the
meaning of probability for individual systems.  As I noted at the
outset, the strategy of the Ithaca interpretation is to set aside such
questions, not because they are unimportant, but because the
interpretation of quantum mechanics has enough problems of its own.
My aim is to find a satisfactory interpretation of quantum mechanics
contingent upon finding a satisfactory understanding of objective
probability as a property of individual systems.  I would consider
that progress.

The question that cannot be evaded, however, is {\it correlations between
what?\/}  I claim that the failure explicitly to formulate and address
this question or to give it only partial answers, is responsible
for many of the most notorious difficulties and anthropomorphisms of
the Copenhagen interpretation: the claim that the existence of a
classical domain is essential for a proper formulation of quantum
mechanics; the intrusion at a fundamental level of notions like
observation, measurement, or state preparation, into what ought to be a
description of phenomena in the unobserved, unmeasured, unprepared
natural world; and the murkiness of the distinction between objective
fact and human knowledge.  

To see how this comes about, note that if correlations between
subsystems of a closed system are indeed the only proper subjects for
physics then the simplest closed non-trivial quantum mechanical system
is not a two-state system, but a four-state system, for a two-state
(or three-state) system cannot describe two non-trivial subsystems.
What is real and objective about such a four-state universe are only
the correlations that exist between the pair of two-state subsystems
it contains.  Observables of one subsystem have no inherent meaning.
They acquire such meaning as they have only from the character of their
correlations with observables of the other subsystem.  If the entire
universe consistented of a two-site spin-$\fr1/2$ Heisenberg model the
complete objective facts about that universe would be subsumed by the
density matrix of that Heisenberg model --- i.e. by nothing more or
less than the collection of all the correlations between the two
subsystems.  To ask about the {\it nature\/} of the correlated
quantities is to go outside of the universe, for it can only be to ask
how they are correlated with something else, and in this toy universe
there {\it is\/} nothing else.

And that's all there is to it for a pair of two-state systems.\ft{See
Appendix B for some of the requirements even so simple a system
imposes on the character of objective probabilities.} Other toy
universes are, of course, more complicated, but what is real and
objective about them is nothing more or less than all the correlations
among their subsystems.  What's real about {\it the\/} Universe (if
you insist on talking about {\it the\/} Universe) are the correlations
among its subsystems.

These correlations constitute the totality of the internal objective
reality of individual systems.  So what do measurement, or a classical
domain, or knowledge have to do with objective reality?  Nothing ---
nothing whatever.  They have to do with {\it us\/}.

We're big complicated systems, and we've evolved under the pressure of
having to deal with other big complicated systems.  We understand
them, we can apprehend them, and we've developed language, to
represent them to ourselves or to help us tell each other about them.
But we did not evolve having to deal with simple two level systems or
even complicated atoms.  So the only way we can cope with such
systems, which evolution did not outfit us to apprehend directly, is
to arrange for them to be subsets of larger systems containing
subsystems of the kind we do know something about dealing with.  We
can then learn about the objectively real correlations that exist
between the small and the big subsystems, and try to infer the nature
of the systems inaccessible to our intuition from how they correlate
with the systems we're equipped to deal with.  The larger systems are
called ``classical'', and the process of arranging to correlate them
with the smaller systems is called ``the measurement process''.

In the measurement process as I've just described it, we ourselves
play the role of God, outside of the universe and directly perceiving
these informative correlations.  It's really not like that, of
course.  To put the point more accurately it's necessary to
acknowledge that we ourselves are physical systems, and what actually
emerges from a measurement are the tripartite correlations between us,
the classical subsystem, and the inaccessible subsystem.  It is
because {\it we\/} have developed the ability to make sense of some of
the correlations between ourselves and classical systems, that we get
something useful out of this process.  But this is a property of {\it
us\/} --- not of the inanimate physical world.  Measurement, the
classical world, and human knowledge enter the picture only when we
ask how {\it we\/} can extract information about the correlations that
constitute the world.  The correlations themselves, however, are there
whether or not we take the trouble to learn about them.

The question of {\it how\/} we are able to understand correlations
between ourselves and the accessible ``classical'' systems we have
arranged to correlate with the inaccessible ``quantum'' systems is
known as the problem of consciousness.  It's a very difficult problem
--- much more difficult, in my opinion, than the interpretation of
quantum mechanics.  But it is a problem about {\it us}.  It is not a problem
that has anything to do with what is objectively real about those
parts of the physical world that can be well isolated from us.

\bigskip

If the first pillar of the Ithaca Interpretation is that correlations
are the only fundamental and objective properties of the world, the
second is that the density matrix of a system is a fundamental
objective property of that system whether or not it is a
one-dimensional projection operator.  To put it another way, in a
nomenclature almost designed to obscure the point, ``mixed'' states
are as fundamental as ``pure'' states.  This flies in the face of much
textbook talk about density matrices.

The problem, of course, is that density matrices can serve two
purposes.  One may indeed be dealing with an ensemble of isolated
systems, each of which has a one-dimensional projection operator as
its density matrix, and want to average over the ensemble the internal
correlations that prevail in each of the subsystems.  The mathematical
object you need to do this has exactly the same structure, but not at
all the same significance, as the fundamental irreducible density
matrix of an individual system.  It is the latter density matrix that
fully describes all the internal correlations of one of the members of
a {\it single\/} EPR pair.  

It remains to be seen whether this point of view toward density
matrices can be developed without running into trouble.  It will be
important that the development of the Ithaca interpretation must be in
a framework that makes it possible to formulate everything entirely in
terms of internal correlations of isolated individual systems. My
guess is that this will be enough to make everything work.  Certain
common but obscure statements about pure vs.~mixed states already make
straightforward sense in this new framework.  For example it is often
said that the difference between a pure state and a mixed state is
that in the former case ``we'' have maximal ``knowledge'' about the
system, while in the latter case ``we'' do not ``know'' everything
that can be ``known''.  The anthropomorphisms disappear
completely if one states this in terms of correlations between
subsystems:

The density matrix of a subsystem $\s_1$ can be a one-dimensional
projection operator (i.e. a pure state) if and only if the only larger
systems $\s = \s_1\+\s_2$ that can contain $\s_1$ as a subsystem admit
of no correlations whatever between $\s_1$ and $\s_2$.  The absence of
such correlations is the objective fact.  The anthropomorphisms simply
express the consequences of this fact for us, should we wish to learn
about $\s_1$.

It is the program of the Ithaca interpretation to reduce all
``quantum mysteries and horrors'' to such statements about objective
probabilities of individual systems.

\bigskip

By not making it explicit that the pure state of a system (when it has
one --- and the density matrix, when it does not) is nothing more than
a concise way to summarize and reveal the consistency of all the
correlations among its subsystems, the Copenhagen interpretation
leaves a conceptual vacuum that is often filled with the implicit and
sometimes explicit notion that its pure quantum state is a fundamental
and irreducible property of a system under study, or even of the
entire world.  By conferring physical reality on the quantum state one
creates a major part of the quantum measurement problem.  I am not
claiming at this point that granting reality only to correlations
among subsystems solves the measurement problem, but it certainly
makes it harder to state just what the problem is.  Because everything
you can formulate in terms of state vectors can also be stated
entirely in terms of correlations between subsystems --- i.e.  in
terms of probability distributions --- if a quantum measurement
problem remains it is going to be a problem about the nature of
objective probabilities of individual systems.  

It is my optimistic expectation that by making the effort to
reformulate the ``measurement problem'' in those terms one will either
demonstrate that it has vanished, or learn something new and important
about the nature of objective probability.

\bigskip
\centerline{{\bf Acknowledgments}}\nobreak\medskip\nobreak I first
encountered the view that correlations are fundamental and irreducible
when I heard it advocated as the proper way to think about
Einstein--Podolsky--Rosen correlations, in talks by Paul Teller and
Arthur Fine.\ft{Published in {\it Philosophical Consequences of
Quantum Theory\/}, James T. Cushing and Ernan McMullin, eds., Notre
Dame Press, Notre Dame, Indiana, 1989.} It did not then occur to me
that this might be the proper way to think about much more general
correlations.\ft{But it should have.  This seems to be the point of
Bohr's reply to EPR: namely, that there is nothing new or unusual
about EPR correlations; precisely the same kinds of correlations are
set up in the measurement process, and therefore there is no cause for
alarm because he, Bohr, has already straightened that out.} Nor did it
occur to me that objective reality might consist {\it only\/} of
correlations until I heard Lee Smolin\ft{Lecture in Bielefeld Germany,
August 1995 (unpublished).} sketch an approach to quantum mechanics
that treated {\it symmetrically\/} a physical system and the world
external to that physical system.  Shortly thereafter I received a
paper from Carlo Rovelli,\ft{ To appear in International Journal of
Modern Physics, {\bf 35}, No. 8 (1996).  See also xxx.lanl.gov e-Print
archive, quant-ph/9609002.} arguing from a very different point of
view that quantum states were nothing more than expressions of
relations between subsystems.  A similar point of view toward quantum
states goes at least back to Everett's original ``relative--state''
formulation of quantum mechanics\ft{Hugh Everett, III, Revs. Mod.
Phys. {\bf 29}, 454-462 (1957).  See also the gloss on Everett in the
paper by John Wheeler that follows, 151-153.} before it was swept off
into the many--worlds extravaganza.\ft{Many worlds might threaten
practitioners of the Ithaca interpretation when they've cleared up the
problems of quantum mechanics and start worrying about how to
understand objective probability, but I trust they will by then be
wise enough to avoid them.} I acquired the notion that certain density
matrices were just as fundamental and irreducible as pure states from
Rudolf Peierls, who insisted to me several years ago that the proper
conclusion to draw from EPR was not non-locality, but the absence of
any objective difference between mixtures of photons with random 0-90
degree polarizations, or random 45-135 degree polarizations.  After
the Bielefeld conference I had an instructive e-mail argument
with Tim Maudlin about this point, and about some analysis by Sandu
Popescu\ft{Physical Review Letters {\bf 74}, 2619-2622 (1995).} that
confirmed my growing suspicion that conventional views about density
matrices and ``quantum non-locality'' were inadequate.  I have also
learned from David Albert, John Bell, Rob Clifton, J\"org Dr\"ager,
Anupam Garg, Gian-Carlo Ghirardi, Nicholas Gisin, Kurt Gottfried, Dan
Greenberger, Bob Griffiths, Lucien Hardy, Jon Jarrett, Thomas Jordan,
Tony Leggett, Yuri Orlov, Phillip Pearle, Asher Peres, Oreste
Piccioni, Abner Shimony, and Henry Stapp.  This work has been
supported by the National Science Foundation, Grant No. PHY-9320831.
\vfil\eject
\centerline{{\bf Appendix A: Remote Construction of Arbitrary
Ensembles}} \centerline{{\bf With a Given Density Matrix}} \bigskip

Any density matrix $W$ is hermitian and can therefore be
expressed in terms of the orthonormal (but not necessarily complete)
set $\k{\phi_i}$ of its eigenvectors with non-zero eigenvalues: $$W =
\sum_{i=1}^d p_i\k{\phi_i}\b{\phi_i}
\eq(Wphi)$$ (with all $p_i > 0$.)  There are alternative ways to
interpret $W$ as distributions of pure states, each of the form: $$W =
\sum_{\mu=1}^D q_\mu\k{\psi_\mu}\b{\psi_\mu},
\eq(Wpsi)$$ where $D
\geq d$, and the (normalized) states $\k{\psi_\mu}$ are not in general
orthogonal.  

The $\k{\psi_\mu}$ must span the same space as the $\k{\phi_i}$, since
the spaces spanned by either set have an orthogonal complement which
is just the set of all $\k{\chi}$ with $\b{\chi}W\k{\chi} = 0$.

Consequently there is an expansion
$$\sqrt{q_\mu}\,\k{\psi_\mu} = \sum_{i = 1}^d M_{\mu i} 
\sqrt{p_i}\,\k{\phi_i}.\eq(M)$$ Because the $\k{\phi_i}$ are an
orthonormal set, for \(Wphi) and \(Wpsi) to yield the
same density matrix $W$ we must have
$$\sum_{\mu=1}^D M_{\mu i}M^*_{\mu j} = \delta_{ij}.\eq(Morth)$$

If $D > d$ we can extend $M$ to a $D$-dimensional unitary
matrix\ft{This is simply the assertion that $d$ orthonormal complex
$D$-vectors $M_{\mu1},\ldots,M_{\mu d}$ can be extended to an
orthonormal basis  $U_{\mu1},\ldots,U_{\mu D}$ for the entire $D$-dimensional space.} $U$ with
$$U_{\mu\nu} = M_{\mu\nu}\,\,,\ \ \ \nu \leq d.\eq(extend)$$
  It follows
from \(M) and the unitarity of $U$ that $$ \sum_{\mu=1}^D
U^*_{\mu\nu}\sqrt{q_\mu}\,\k{\psi_\mu} = 0,\ \ \ \nu > d.
\eq(vanish)$$

We now define a state in the product of our original state space and a
space of dimension $D$: $$\k{\Phi} = \sum_{i=1}^d
\sqrt{p_i}\,\k{\phi_i}\x\k{\alpha_i},\eq(epr)$$ where the
$\k{\alpha_i}$ are the first $d$ members of an (arbitrarily chosen)
orthonormal set $\k{\alpha_\mu}$,\ $\mu=1\ldots D$.

It follows from \(M) and \(Morth) that 
$$\sqrt{p_i}\,\k{\phi_i} = \sum_{\mu=1}^D \sqrt{q_\mu}\,\k{\psi_\mu}M^*_{\mu
i}\,\,,\eq(Minv)$$ and therefore
$$\k{\Phi} = \sum_{\mu=1}^D\sqrt{q_\mu}\,\k{\psi_\mu} \sum_{i=1}^d M^*_{\mu i}
\x\k{\alpha_i}. \eq(epr1)$$ Eq. \(vanish) permits us to extend the
sum to the entire set of $D$ vectors $\k{\alpha_\mu}$:  $$\k{\Phi} =
\sum_{\mu=1}^D\sqrt{q_\mu}\,\k{\psi_\mu} \sum_{\nu=1}^D U^*_{\mu \nu}
\x\k{\alpha_\nu}. \eq(epr2)$$

We have thus arrived at an alternative form
$$\k{\Phi} = \sum_{\mu=1}^D\sqrt{q_\mu}\,\k{\psi_\mu}
\x\k{\beta_\mu},\eq(epr3)$$
where $$\k{\beta_\mu} = \sum_{\nu=1}^D
U^*_{\mu \nu}
\k{\alpha_\nu}. \eq(epr1)$$ It follows from the unitarity of $U$
and the orthonormality of the $\k{\alpha_\mu}$ that the
$\k{\beta_\mu}$ are also an orthonormal set.

If we are given a large number of alternative realizations of $W$ of
the form \(Wpsi), we can take the dimension of the auxilliary space to
be the largest $D$ associated with them.  The above argument then
shows that if we are given any state $\k{\Phi}$ of the form \(epr), we
can find a representation of $\k{\Phi}$ having the form \(epr3) for
any of the many sets of $\k{\psi_\mu}$ satisfying \(Wpsi).  By
measuring in the auxilliary space an observable whose eigenstates are
the associated $\k{\beta_\mu}$, we can therefore produce an ensemble
in the original space in which the system is in the state
$\k{\psi_\mu}$ with probability $q_\mu$.

\bigskip \centerline{{\bf Appendix B: The Hardy Paradox.}}  \nobreak
\medskip \nobreak The simplest possible non-trivial closed individual
quantum system --- a pair of two two-state systems --- already gives
some useful clues about some of the properties objective probabilities
will have to possess.  The following example, invented by Lucien Hardy
to give a particularly powerful version of Bell's Theorem, also
enables one to make an important point about objective probabilities.

Call the two two-state subsystems $A$ and $B$.  To make the point we
need consider only two observables of each system, called $1_A$,
$2_A$, $1_B$, and $2_B$.  We can label the two eigenstates of each of
these observers by a color: red ($R$) or green ($G$).  In each
subsystem take the eigenstates of observable 1 to be non-trivially
different from those of observable 2 --- i.e. $\k{1R}$ is a
superposition of $\k{2R}$ and $\k{2G}$ with both coefficients
non-zero.  To make the point it suffices to take the symmetric case in
which the values of the two coefficients are the same, whether the
observables $1$ and $2$ are associated with subsystem $A$ or subsystem
$B$.  To keep the notation from getting too cumbersome we abbreviate
the designation of a state of the form $\k{1_AR}\x\k{2_BG}$ (for
example) simply to $\k{1R,2G}$.

Now consider the universe consisting of the pair of two-state systems
characterized by the density matrix $\k\Psi\b\Psi$ which projects on the
(normalized) state:
$$\k{\Psi} = {\k{2R,2R} - \k{1R,1R}\ip{1R,1R}{2R,2R} 
    \over \sqrt{1 - \ip{1R}{2R}^4}}.\eq(hardy)$$ 
Clearly $$ p(1R,1R) = |\ip{1R,1R}{\Psi}|^2 = 0, \eq(5)$$
        $$ p(2G,1G) = |\ip{2G,1G}{\Psi}|^2 = 0, \eq(6)$$
        $$ p(1G,2G) = |\ip{1G,2G}{\Psi}|^2 = 0, \eq(7)$$ 
while  $$ p(2G,2G) = |\ip{2G,2G}{\Psi}|^2 =  
{(1-x)^2x^2 \over 1-x^2} = x^2\Bigl({1-x \over 1+x}\Bigr),
\eq(8)$$
where $$ x =
|\ip{1R}{2R}|^2 \neq 0. \eq(2)$$

The only important thing to note is that the first three of these
probabilities are zero and the fourth is non-zero, but I cannot resist
noting that the probability $p(2G,2G)$ happens to be maximum when $x =
1/\tau$ (where $\tau$ is the golden mean, $\tau = {\sqrt5 + 1 \over
2}$), in which case the values of all the probabilities associated with
the four pairs of subsystem observables are as in the following lovely Table:

\bigskip
\def\tt{\thinspace\thinspace}
\def\t#1{$\tau^{-#1}$}
\bigskip
\centerline{
\setbox\strutbox=\hbox{\vrule height12pt depth6.5pt width0pt}
\vbox{\offinterlineskip
\hrule
\halign{
         \strut\vrule\hfil\thinspace #\thinspace\hfil
        &\vrule\vrule\vrule\hfil\tt #\thinspace\hfil
        &\vrule\hfil\tt #\thinspace\hfil      
        &\vrule\hfil\tt #\tt\hfil 
        &\vrule\hfil\tt # \hfil\vrule\cr 
$p$ & 22  & 11  & 12  & 21 \cr
\noalign{\hrule}
\noalign{\hrule}
\noalign{\hrule}
\noalign{\hrule}
GG & \t5 & \t3 & 0 & 0 \cr
\noalign{\hrule}
GR & \t4 & \t2 & \t1 & \t3  \cr
\noalign{\hrule}
RG & \t4 & \t2 & \t3  & \t1 \cr
\noalign{\hrule}
RR & \t1 &  0  & \t4  & \t4 \cr
\noalign{\hrule}}}}
\bigskip
The Hardy paradox consists of observing that the three 0 probabilities
translate into three conditional probabilities of unity: 

$$p(1_AG,2_BG) = 0 \imp p(2_BG) = p(1_AR,2_BG) \imp p(1_AR|2_BG) =
1,\eq(9a)$$ $$p(1_AR,1_BR) = 0 \imp p(1_AR) = p(1_AR,1_BG) \imp
p(1_BG|1_AR) = 1,\eq(9b)$$ $$p(2_AG,1_BG) = 0 \imp p(1_BG) =
p(1_BG,2_AR) \imp p(2_AR|1_BG) = 1.\eq(9c)$$ From these unit
conditional probabilities we conclude that $2_BG$ requires $1_AR$,
that $1_AR$ requires $1_BG$, and that $1_BG$ requires $2_AR$.
Therefore $2_BG$ requires $2_AR$: $$ p(2_AR|2_BG) = 1.\eq(contra)$$
But this contradicts the fact \(8) that $$p(2_AG,2_BG) \neq
0.\eq(contra1)$$

The conventional analysis of what's  wrong with this reasoning associates
the probabilities with the results of measurements.  Thus the
probability $$ p(1_AR|2_BG) = 1\eq(10)$$ appearing in
\(9a) must actually be conditioned not only on getting $G$ for a
measurement of $2_B$, but also on both measurements actually being
performed.  We should therefore use the expanded form $$ p(1_AR|2_BG;
1_A,2_B) = 1.\eq(11)$$ The second $2_B$ is unnecessary, if we
interpret $2_BG$ to mean {\it property $2_B$ is measured and found to
have the value $G$}.  It might appear that the second $1_A$ is also
unnecessary, but this is incorrect.  For the naive argument to go
through, the $1_AR$ in \(9a) must mean exactly the
same thing as it means in \(9b) --- namely, {\it property
$1_A$ is measured and found to have the value $R$}.  But the
probability is not 1 that if $2_B$ is measured and found to have the
value $G$ then $1_A$ is measured and found to have the value $R$.  To
get a probability of 1 we must also condition on subsystem $1_A$ actually
being measured.  Therefore we must rewrite \(9a)-- \(9c)
as $$\eqalign{ p(1_AR|2_BG; 1_A) &= 1, \cr p(1_BG|1_AR; 1_B) &= 1, \cr
p(2_AR|1_BG; 2_A) &= 1, \cr }\eq(12)$$ and the chain of reasoning
following \(9c) breaks down.

This way out of Hardy's paradox is not available to the Ithaca
interpretation, which insists that quantum mechanics should make sense
as a description of the objectively real correlations that exist in a
universe consisting {\it entirely\/} of the two two-state systems.  In
such a universe there are no measurements --- only correlations.  The
additional conditioning on an observable ``actually being measured''
has no meaning.  In the Ithaca interpretation the fallacy in the Hardy
paradox can only be that the three ``conditional probabilities'' equal
to unity in \(9a)-\(9c) have no meaning.  It makes no sense to
contemplate the probability that $1_A$ is $R$ {\it given\/} that $2_B$
is $G$.  The unconditional value of an observable for a subsystem
cannot be ``given'' --- only correlations between subsystems have
objective reality.

It therefore appears that the view of probability underlying the
Ithaca interpretation must be anti-Bayesian.  At some fundamental
level unconditional joint objective probabilities have meaning, but
certain conditional probabilities have no meaning, because that
upon which they are conditioned has no objective reality.  Only
correlations --- i.e. only joint distributions --- have objective
reality.

\bye